\documentstyle[12pt]{article}

\def\hybrid{\topmargin -20pt	\oddsidemargin 0pt
	\headheight 0pt	\headsep 0pt
	\textwidth 6.25in	
	\textheight 9.5in	
	\marginparwidth .875in
	\parskip 5pt plus 1pt	\jot = 1.5ex}

\hybrid

\def\baselinestretch{1.2}

\catcode`\@=11

\def\marginnote#1{}
\newcount\hour
\newcount\minute
\newtoks\amorpm
\hour=\time\divide\hour by60
\minute=\time{\multiply\hour by60 \global\advance\minute by-\hour}
\edef\standardtime{{\ifnum\hour<12 \global\amorpm={am}%
	\else\global\amorpm={pm}\advance\hour by-12 \fi
	\ifnum\hour=0 \hour=12 \fi
	\number\hour:\ifnum\minute<10 0\fi\number\minute\the\amorpm}}
\edef\militarytime{\number\hour:\ifnum\minute<10 0\fi\number\minute}

\def\draftlabel#1{{\@bsphack\if@filesw {\let\thepage\relax
   \xdef\@gtempa{\write\@auxout{\string
      \newlabel{#1}{{\@currentlabel}{\thepage}}}}}\@gtempa
   \if@nobreak \ifvmode\nobreak\fi\fi\fi\@esphack}
	\gdef\@eqnlabel{#1}}
\def\@eqnlabel{}
\def\@vacuum{}
\def\draftmarginnote#1{\marginpar{\raggedright\scriptsize\tt#1}}

\def\draft{\oddsidemargin -.5truein
	\def\@oddfoot{\sl preliminary draft \hfil
	\rm\thepage\hfil\sl\today\quad\militarytime}
	\let\@evenfoot\@oddfoot	\overfullrule 3pt
	\let\label=\draftlabel
	\let\marginnote=\draftmarginnote
   \def\@eqnnum{(\theequation)\rlap{\kern\marginparsep\tt\@eqnlabel}%
\global\let\@eqnlabel\@vacuum}  }

\def\preprint{\twocolumn\sloppy\flushbottom\parindent 2em
	\leftmargini 2em\leftmarginv .5em\leftmarginvi .5em
	\oddsidemargin -.5in	\evensidemargin -.5in
	\columnsep .4in	\footheight 0pt
	\textwidth 10.in	\topmargin  -.4in
	\headheight 12pt \topskip .4in
	\textheight 6.9in \footskip 0pt
	\def\@oddhead{\thepage\hfil\addtocounter{page}{1}\thepage}
	\let\@evenhead\@oddhead	\def\@oddfoot{}	\def\@evenfoot{} }

\def\numberbysection{\@addtoreset{equation}{section}
	\def\theequation{\thesection.\arabic{equation}}}

\def\underline#1{\relax\ifmmode\@@underline#1\else
	$\@@underline{\hbox{#1}}$\relax\fi}

\def\titlepage{\@restonecolfalse\if@twocolumn\@restonecoltrue\onecolumn
     \else \newpage \fi \thispagestyle{empty}\c@page\z@
	\def\thefootnote{\fnsymbol{footnote}} }

\def\endtitlepage{\if@restonecol\twocolumn \else \newpage \fi
	\def\thefootnote{\arabic{footnote}}
	\setcounter{footnote}{0}}  

\catcode`@=12
\relax

\def\figcap{\section*{Figure Captions\markboth
	{FIGURECAPTIONS}{FIGURECAPTIONS}}\list
	{Figure \arabic{enumi}:\hfill}{\settowidth\labelwidth{Figure 999:}
	\leftmargin\labelwidth
	\advance\leftmargin\labelsep\usecounter{enumi}}}
 \relax
\def\tablecap{\section*{Table Captions\markboth
	{TABLECAPTIONS}{TABLECAPTIONS}}\list
	{Table \arabic{enumi}:\hfill}{\settowidth\labelwidth{Table 999:}
	\leftmargin\labelwidth
	\advance\leftmargin\labelsep\usecounter{enumi}}}
 \relax
\def\reflist{\section*{References\markboth
	{REFLIST}{REFLIST}}\list
	{[\arabic{enumi}]\hfill}{\settowidth\labelwidth{[999]}
	\leftmargin\labelwidth
	\advance\leftmargin\labelsep\usecounter{enumi}}}
 \relax
\makeatletter
\newcounter{pubctr}
\def\publist{\@ifnextchar[{\@publist}{\@@publist}}
\def\@publist[#1]{\list
	{[\arabic{pubctr}]\hfill}{\settowidth\labelwidth{[999]}
	\leftmargin\labelwidth
	\advance\leftmargin\labelsep
	\@nmbrlisttrue\def\@listctr{pubctr}
	\setcounter{pubctr}{#1}\addtocounter{pubctr}{-1}}}
\def\@@publist{\list
	{[\arabic{pubctr}]\hfill}{\settowidth\labelwidth{[999]}
	\leftmargin\labelwidth
	\advance\leftmargin\labelsep
	\@nmbrlisttrue\def\@listctr{pubctr}}}
 \relax
\makeatother
\newskip\humongous \humongous=0pt plus 1000pt minus 1000pt

\newif\ifdtup

\relax

\def\thefootnote{\fnsymbol{footnote}}
\def\be{\begin{equation}}
\def\ee{\end{equation}}
\def\ba{\begin{eqnarray}}
\def\ea{\end{eqnarray}}

\begin{document}
\renewcommand{\theequation}{\arabic{equation}}
\newcommand{\beq}{\begin{equation}}
\newcommand{\eeq}[1]{\label{#1}\end{equation}}
\newcommand{\ber}{\begin{eqnarray}}
\newcommand{\eer}[1]{\label{#1}\end{eqnarray}}
\begin{titlepage}
\begin{center}

\hfill CERN-TH/96-201\\
\hfill SNUTP 96-79\\
\hfill hep-th/9607243\\

\vskip .5in

{\large \bf Gravitational Dressing of Massive Soliton Theories}

\vskip 0.5in

{\bf Ioannis Bakas}
\footnote{Permanent address: Department of Physics, University of Patras, 
26110 Patras, Greece}
\footnote{e-mail address: BAKAS@NXTH04.CERN.CH}\\
\vskip .1in

{\em Theory Division, CERN, 1211 Geneva 23, Switzerland}\\ 

\vskip .4in

{\bf Q--Han Park}
\footnote{e-mail address: QPARK@NMS.KYUNGHEE.AC.KR}\\

\vskip .1in

{\em Department of Physics, and Research Institute of Basic Sciences\\
Kyunghee University, Seoul, 130-701, Korea}\\
\end{center}

\vskip .9in

\begin{center} {\bf ABSTRACT } \end{center}
\begin{quotation}\noindent
The massive soliton theories describe integrable perturbations
of WZW cosets as generalized multi-component sine-Gordon models.
We study their coupling to 2-dim gravity in the conformal gauge
and show that the resulting models can be interpreted as
conformal non-Abelian Toda theories when a certain algebraic
condition is satisfied. These models, however, do not provide 
quantum mechanically consistent string backgrounds in the case 
the underlying WZW constraints are first solved classically.
\end{quotation}
\vskip1.5cm
July 1996\\
\end{titlepage}
\vfill
\eject

\def\baselinestretch{1.2}
\baselineskip 16 pt
\setcounter{equation}{0}
\noindent
It is well known that any Lagrangian can be made scale 
invariant by appropriate coupling to a bosonic 
field $\phi$ that transforms non-linearly 
$\phi (x) \rightarrow \phi (e^{\alpha} x) + \alpha / \gamma$,
where $\gamma$ is a constant and $\alpha$ is the scale 
parameter. Let $L = L_{s} - V$ be the 
decomposition of the original Lagrangian into a scale 
invariant part $L_{s}$ and a scale breaking potential $V$. 
Then, 
\be
L^{\prime} = L_{s} - e^{2 \gamma \phi} V + 
{1 \over 2 {\gamma}^2} {\partial}_{\mu} e^{\gamma \phi} 
{\partial}^{\mu} e^{\gamma \phi} 
\ee
is manifestly scale invariant and also conformal. This 
prescription was used in the past to write down 4-dim
phenomenological Lagrangians for which mesons and nucleons
have non-zero mass in lowest order perturbation theory 
and only $\phi$ is massless serving in this case 
as the Goldstone boson of scale 
invariance (see for instance [1] and references therein).

The above procedure has a natural interpretation in 
two dimensions by providing the coupling of a relativistic
theory to gravity in the conformal gauge. 
For instance, we may
think of $L$ as an integrable perturbation of a 
conformal field theory, typically given by a gauged WZW
Lagrangian $L_{s}$ plus a potential term $V$ that breaks
conformal invariance while preserving integrability 
\`a la Zamolodchikov [2]. The coupling to 2-dim gravity 
amounts to choosing a metric $g = e^{2 \gamma \phi} 
\widehat{g}$ with the corresponding action becoming
\be
S^{\prime} = S_{s} - \int d^2 z \sqrt{\widehat{g}} ~
e^{2 \gamma \phi} V + S_{L} ~,
\ee
where the contribution of the Liouville field $\phi$ is 
\be
S_{L} = {1 \over \pi} 
\int d^2 z \sqrt{\widehat{g}} \left( {1 \over 2} 
{\widehat{g}}^{ab} {\partial}_{a} \phi {\partial}_{b} 
\phi + Q \phi R(\widehat{g}) + 
\mu e^{2 \gamma \phi} \right) .
\ee
The contribution of the dilaton term vanishes by 
specializing to flat ${\widehat{g}}_{ab}$, while the 
conformal factor transforms as usual under 
$z \rightarrow f(z)$,
\be
\phi (z) \rightarrow \phi (f(z)) + {1 \over 2 \gamma} \log 
{\mid \partial f \mid}^2 ~, 
\ee 
in terms of light-cone variables. 
The cosmological term $\mu$ arises
due to the freedom to shift $V$ by a  
constant that changes the zero point energy without 
affecting the classical equations of motion of the system
before coupling it to gravity. This is the general framework
of the gravitational dressing; see the original papers [3]
for further details.

There are already some results in the literature on the 
gravitational sine-Gordon model, which arises by perturbing 
the free field action by a cosine potential and then coupling
it to the Liouville theory as above 
[4, 5]. Here we consider the
more general situation of a 2-dim relativistic system 
described by a conformal coset model $G/H$, plus an 
integrable perturbation, and determine the circumstances
under which the coupling to gravity preserves the 
integrability at least classically. The most typical
example in our scheme, besides the simplest sine-Gordon
model, is the so called complex 
sine-Gordon (here we actually consider its analytic 
continuation with hyperbolic elements) with Lagrangian
\be
L = \partial r \bar{\partial} r - {\tanh}^2 r 
\partial \theta \bar{\partial} \theta + 
\lambda \cosh 2r
\ee
with coupling constant $\lambda$. 
This model was recently interpreted as an integrable
perturbation of the WZW coset $SL(2)/U(1)$ by the first
thermal (parafermionic) operator ${\epsilon}_1$ that 
turns on the potential term $\cosh 2r$ [6, 7]; 
similar results hold for the compact coset $SU(2)/U(1)$. 
Upon gravitational dressing, according to the general 
framework, the resulting conformal theory for 
appropriately chosen constants 
$\gamma$ and $\lambda$ assumes the form
\be
L^{\prime} = \partial r \bar{\partial} r - {\tanh}^2 r
\partial \theta \bar{\partial} \theta + \partial \phi 
\bar{\partial} \phi + e^{2 \phi} \cosh 2r ~, 
\ee
which is easily recognized as the non-Abelian $B_2$ Toda
model [8, 9]. Non-Abelian Toda theories will be the center 
theme of our study later. They were also investigated before 
as exactly solvable conformal systems of the matter field 
$\phi$ 
in the presence of black-hole backgrounds (the simplest 
example being the semi-classical geometry of the coset
$SL(2)/U(1)$) [9], but their interpretation in terms of 
the gravitational dressing of integrable perturbations 
of coset models appears to be new in the present context. 

Before proceeding further we note for completeness that
apart from the ordinary sine-Gordon model various aspects 
of some other gravitationally dressed relativistic 
integrable systems have also been considered elsewhere:
for the Gross-Neveu model coupling to gravity 
in the chiral (Polyakov) gauge
[10], and for the principal chiral model in the 
conformal gauge [11]. The theories we consider 
below form a different class that is rather wide. 
A primary aim is to accumulate 
further results in the developing area of gravitationally 
dressed systems by encompassing integrable models 
like (5), (6), 
and multi-component generalizations thereof, 
thus establishing new connections among them in terms 
of 2-dim gravity.

The general framework of the present work is given by 
the class of massive integrable soliton theories (MIST),
as was formulated in [12], and their algebraic connection
with non-Abelian Toda theories. Recall first the essential 
technical ingredients that define the relevant class of
integrable models. We consider the $sl(2)$ embedding of 
a finite algebra ${\bf g}$ specified by the generators 
$\{J_{\pm} , J_{0}\}$ with
\be
[J_{0} , ~ J_{\pm}] = \pm J_{\pm} ~, ~~~~~ 
[J_{+} , ~ J_{-}] = 2 J_{0} ~,
\ee
so that the Cartan element $J_{0}$ induces a gradation
of ${\bf g}$ as follows:
\be
{\bf g} = {\bf g}_{-1} \oplus {\bf g}_{0} \oplus 
{\bf g}_{1} ~; ~~~~~
[J_{0} , ~ {\bf g}_{k}] = k {\bf g}_{k} ~, ~~~
k = 0, ~\pm 1 ~.
\ee
In this case we speak of an $N=1$ grading of ${\bf g}$ 
and ${\bf g}_{0}$,
which is the zero graded part of ${\bf g}$, 
forms a subalgebra. More general gradations correspond
to the decomposition ${\bf g} = \oplus_{k= -N}^{k= N}
{\bf g}_k$, but these will not enter at all into the
present work.
Let $G_{0}$ denote the Lie group associated with 
${\bf g}_{0}$ and write down the action
\be
S = S_{WZW} (g, A, \bar{A}) + {m^2 \over 2 \pi} 
\int {\rm Tr} (g^{-1} T g \bar{T}) ~,
\ee
where $g$ is an element of $G_{0}$ and $T$, $\bar{T}$ are
constant elements of the Lie algebra ${\bf g}$ (though they
remain arbitrary at the moment). 
The term $S_{WZW}$ is the usual gauged WZW action 
$(1/4 \pi) \int {\rm Tr} (g^{-1} \partial g g^{-1} 
\bar{\partial} g) + WZ{\rm -term} + 
(A, ~ \bar{A}){\rm -terms}$, where ${\rm Tr}(T^{a}T^{b})
= 2 {\delta}^{ab}$.
The gauge connections $A$, $\bar{A}$
take values in the subgroup $H$ of $G_{0}$ whose algebra
is defined by
\be
{\bf h} = \{v \in {\bf g}_{0} ; ~~ [v, ~ T] = 0 = 
[v, ~ \bar{T}] \} ~.
\ee
This on the other hand specifies the flat directions of 
the potential in (9), since the potential term is invariant
under the adjoint action of $H$ associated with the Lie
algebra ${\bf h}$. 

For the WZW model $G_{0} / H$, where $H$ is assumed to be 
diagonal subgroup of $G_{0}$, $A$ and $\bar{A}$ act as 
Lagrange multipliers resulting in the constraint equations
\ba
{\delta}_{A} S & = & {1 \over 2 \pi} \int {\rm Tr} 
(- \bar{\partial}g g^{-1} + g \bar{A} g^{-1} - \bar{A}) 
\delta A = 0 ,\\
{\delta}_{\bar{A}} S & = & {1 \over 2 \pi} \int {\rm Tr}
(g^{-1} \partial g + g^{-1} A g - A) \delta \bar{A} = 0
\ea
that remove the flat directions of the potential. The 
remaining equations of motion take the form
\be
[\bar{\partial} + \bar{A} , ~ g^{-1} D g] = 
m^2 [T, ~ g^{-1} \bar{T} g]
\ee
or in zero curvature form they become
\be
[\partial + g^{-1} D g + l T , ~ \bar{\partial} + \bar{A} 
+ {m^2 \over l} g^{-1} \bar{T} g] = 0 ~,
\ee
where $g^{-1} D g = g^{-1} \partial g + g^{-1} A g$ and $l$
is a spectral parameter. Projecting (13) in ${\bf h}$ and using
the constraint equations it follows that the gauge field is
flat, $[\partial + A , ~ \bar{\partial} + \bar{A}] = 0$, which
in turn allows for the particularly simple gauge fixing
\be
A = 0 = \bar{A} ~.
\ee
We then arrive at the equations
\be
\bar{\partial} (g^{-1} \partial g) = m^2 [T , ~ g^{-1} 
\bar{T} g] ~ ; ~~~~~ (g^{-1} \partial g)_{{\bf h}} = 0 
= (\bar{\partial} g g^{-1})_{{\bf h}}
\ee
taking into account the projection in ${\bf h}$ as it is 
explicitly indicated in (16). 

Within the above general framework we may specialize the 
class of integrable field theories (9) according to 
specific choices for $T$ and $\bar{T}$. Here we will 
consider two cases:

\noindent
(a) {\em Massive models}, choosing 
$T = \bar{T} = J_{+} + J_{-}$
for which the potential explicitly breaks conformal 
invariance acting as a mass term. We obtain in this case
soliton equations that provide a non-Abelian extension of
the sine-Gordon model to soliton theories with internal
degrees of freedom [12, 13]. The complete list of the
massive integrable soliton theories associated to $N=1$
grading of all classical Lie algebras ${\bf g}$ (MIST)
has been obtained in [12]. These models are invariant 
under parity.

\noindent
(b) {\em Conformal models}, choosing $T = J_{+}$ and 
$\bar{T} = J_{-}$. In this case the potential term 
possesses the right scaling properties for defining 
conformally invariant models known as non-Abelian 
Toda theories [8, 9] generalizing the Liouville theory. 

\noindent
In both cases above the non-Abelian nature of the models
originates from the non-Abelian algebra ${\bf g}_{0}$ in 
the $N=1$ grading (8) of ${\bf g}$. The affine 
(non-conformal) Toda theories correspond to the choice 
$T = J_{+} + Y_{+}$ and $\bar{T} = J_{-} + Y_{-}$, where
$Y_{\pm}$ are elements of the maximally graded parts 
${\bf g}_{\mp N}$ for a general gradation. Hence, the
models (a) are specified by the choice $Y_{\pm} = J_{\mp}$.

A crucial property inherited from this algebraic 
model building is the factorization of $G_{0}$ into
a direct product of an abelian $U(1)$ 
factor generated by
$J_{0}$ of $sl(2)$ and the remaining (non-abelian) 
group denoted by $G_{0}^{\prime}$. This decomposition
is the starting point for further refining the class
of MIST under consideration. In fact we will consider 
next those models for which the field $\chi$ 
associated with the abelian $U(1)$ factor can be 
decoupled consistently. We will derive the algebraic
condition that is necessary for reducing the MIST 
models
by setting $\chi = 0$ while maintaining consistency
with the classical equations of motion. The complex
sine-Gordon model can be obtained precisely in this fashion,
but now within the present framework we have a whole
class of possible massive multi-component 
generalizations as well. Upon 
gravitational dressing, as we will see later, 
this particular class of models
admits a systematic description as conformal models
of type (b), i.e. non-Abelian Toda theories; the
simplest example of such a construction/correspondence
is the transition from (5) to (6), which was considered
before as motivation. This also explains why we have
presented here a unified description of the models 
(a) and (b). 

Let $g = g^{\prime} \exp (i \chi J_{0})$ be the 
decomposition of a group element in $G_{0}$ into 
$G_{0}^{\prime}$ and the commuting $U(1)$ factor
generated by $J_{0}$. Then, the equations of motion
(16) of any MIST with $T = \bar{T} = J_{+} + J_{-}$ 
decompose as follows:
\ba
\bar{\partial} (g^{-1} \partial g) & = & 
\bar{\partial} ({g^{\prime}}^{-1} \partial g^{\prime}) 
+ i \bar{\partial} \partial \chi J_{0} = 
m^2 [J_{+} + J_{-} , ~ g^{-1} (J_{+} + J_{-})g]
\nonumber\\
& = & m^{2} e^{i \chi} [J_{+} , ~ {g^{\prime}}^{-1}
J_{-} g^{\prime}] + m^{2} e^{-i \chi} 
[J_{-} , ~ {g^{\prime}}^{-1} J_{+} g^{\prime}] ~,
\ea
where we used the fact that $g^{\prime}$ commutes with
$J_{0}$. Then, the $U(1)$ part of this equation yields
\ba
i \bar{\partial} \partial \chi & = & 
{m^2 \over {\rm Tr} J_{0}^2} {\rm Tr} J_{0} 
(e^{i \chi} [J_{+} , ~ {g^{\prime}}^{-1} J_{-} 
g^{\prime}] + e^{-i \chi} [J_{-} , ~ 
{g^{\prime}}^{-1} J_{+} g^{\prime}]) \nonumber\\
& = & {m^2 \over {\rm Tr} J_{0}^2} 
{\rm Tr} (e^{i \chi} J_{+} {g^{\prime}}^{-1} 
J_{-} g^{\prime} - e^{-i \chi} J_{-} 
{g^{\prime}}^{-1} J_{+} g^{\prime}) ~.
\ea
Thus, the $U(1)$ factor can be consistently decoupled
by setting $\chi = 0$ without contradicting the classical
equations of motion provided that the particular
MIST model satisfies the condition
\be
{\rm Tr} (J_{+} {g^{\prime}}^{-1} J_{-} g^{\prime})
= {\rm Tr} (J_{-} {g^{\prime}}^{-1} J_{+} g^{\prime}) ~.
\ee
The models that result in this fashion are integrable
having as equations of motion
\be
\bar{\partial} ({g^{\prime}}^{-1} \partial g^{\prime}) 
= m^2 ([J_{+} , ~ {g^{\prime}}^{-1} J_{-} g^{\prime}]
+ [J_{-} , ~ {g^{\prime}}^{-1} J_{+} g^{\prime}]) ~.
\ee

Starting now from any consistently reduced MIST model with
$\chi = 0$, the coupling to 2-dim gravity in the 
conformal gauge $g_{ab} = e^{2 \phi} {\delta}_{ab}$ 
amounts to the conformally invariant theory
\be
S^{\prime} = S_{WZW} (g^{\prime} , A , \bar{A}) + 
{2 \over \pi} \int d^2 z \partial \phi \bar{\partial} \phi + 
{m^2 \over \pi} \int d^2 z e^{2 \phi} {\rm Tr} 
(J_{+} {g^{\prime}}^{-1} J_{-} g^{\prime}) ~,
\ee
where we used the special condition (19) for the original
potential term to write down 
${\rm Tr} ((J_{+} + J_{-}) {g^{\prime}}^{-1} 
(J_{+} + J_{-}) g^{\prime}) = 2 {\rm Tr} 
(J_{+} {g^{\prime}}^{-1} J_{-} g^{\prime})$.
Notice that ${\rm Tr}(J_{+} {g^{\prime}}^{-1} J_{+} 
g^{\prime}) = 0$ using (7) and the commutativity of
$J_{0}$ with $g^{\prime}$; indeed 
${\rm Tr}(J_{+} {g^{\prime}}^{-1} J_{+} g^{\prime}) = 
{\rm Tr}([J_{0} , J_{+}] {g^{\prime}}^{-1} J_{+} 
g^{\prime}) = {\rm Tr}(J_{+} {g^{\prime}}^{-1} 
[J_{+} , J_{0}] g^{\prime}) = - {\rm Tr}(J_{+} 
{g^{\prime}}^{-1} J_{+} g^{\prime})$, and so this term 
vanishes. We may prove similarly that 
${\rm Tr}(J_{-} {g^{\prime}}^{-1} J_{-} g^{\prime}) = 0$.
At this point we introduce group elements
\be
\tilde{g} = g^{\prime} e^{2 \phi J_{0}}
\ee
and note that
\be
e^{2 \phi} {\rm Tr} (J_{+} {g^{\prime}}^{-1} J_{-} 
g^{\prime}) = {\rm Tr} (e^{2 \phi J_{0}} J_{+} 
e^{-2 \phi J_{0}} {g^{\prime}}^{-1} J_{-} g^{\prime})
= {\rm Tr} (J_{+} {\tilde{g}}^{-1} J_{-} \tilde{g})~.
\ee
As a result the conformal theory (21) becomes
\be
S^{\prime} = S_{WZW} (\tilde{g} , A, \bar{A}) + 
{m^2 \over \pi} \int d^{2} z {\rm Tr} (J_{+} 
{\tilde{g}}^{-1} J_{-} \tilde{g}) ~,
\ee
which is immediately recognized as a non-Abelian
Toda theory of type (b) with $T = J_{+}$, 
$\bar{T} = J_{-}$. This result summarizes the effect
of the gravitational dressing on the class of 
integrable models satisfying the algebraic condition (19).
Alternatively, it provides a new geometrical interpretation
to a wide class of conformal non-Abelian Toda models.
It is also interesting to recall 
that the massive soliton theories
before dressing them with gravity describe integrable
perturbations of WZW models away from criticality. 
We will present some examples later.

Note that despite of appearances the field 
$\chi$ (before decoupling it from a massive integrable
soliton theory) and the Liouville field $\phi$ (after
dressing the resulting models with 2-dim gravity) 
play a different role: we may effectively interpret 
$\chi$ as providing a sine-Gordon type coupling, 
hence leading to the massive integrable soliton theories 
with broken conformal invariance, 
as in (17), while $\phi$
is a Liouville coupling restoring the conformal
invariance classically. Apart from this, 
the field $\phi$ satisfies a field 
equation following (21) that renders the choice 
$\phi = 0$ inconsistent; a consistent choice is instead
$\phi \rightarrow - \infty$. However, as far as the
WZW part of the action (24) is concerned, it is obvious
that $S_{WZW} (g, A, \bar{A})$ is essentially the
same as $S_{WZW} (\tilde{g}, A, \bar{A})$, modulo
a formal analytic continuation $\chi \rightarrow 
-2i \phi$ that changes the commuting $U(1)$ factor
of $G_{0}$ into the non-compact form $R$. The potential
terms on the other hand are not equivalent; they agree
only when the group elements take values in 
$G_{0}^{\prime}$, i.e. ${\rm Tr} (J_{+} {g^{\prime}}^{-1} 
J_{-} g^{\prime})$.
Of course, $T$ and $\bar{T}$ 
are identified differently in each case. 
This comparison is useful to
understand intuitively the essence of our results
without willing to threaten them with trivialization.

The resulting non-Abelian Toda theories provide a class
of curved backgrounds for string propagation in the
presence of a tachyon field, which is identified with
the Toda potential. A drawback of these models, in
their formulation obtained by solving
classically the WZW constraints, is that they 
are not conformal quantum
mechanically and therefore fail to be solutions of the 
corresponding $\beta$-function equations; we will 
elaborate more on this issue towards the end of the paper.
Thus, our interest in the classically reduced form of
the models is only from the
point of view of integrable 2-dim field theories and not
for applications in string theory.

Next, we examine for which algebras ${\bf g}$ with
$N=1$ grading the decoupling condition (19) is satisfied,
and also present a few explicit examples illustrating the
general situation. 

(i) {\em A-series}: For the $A$-series the rank of the algebra
${\bf g}$ has to be odd, and so we consider $A_{2n-1}$ for
which there is an $N=1$ embedding of $sl(2)$ specified in the
defining representation by
\be
J_{0} = {1 \over 2} \left(\begin{array}{ccc}
1_{n} & & 0 \\
0    & & -1_{n} \end{array}\right) , ~~~~
J_{+} = \left(\begin{array}{ccc}
0  & & 1_{n}\\
0  & &   0 \end{array}\right) , ~~~~
J_{-} = \left(\begin{array}{ccc}
0  &  & 0\\
1_{n} & & 0 \end{array}\right) . 
\ee
This results in the zero-graded part of the algebra
${\bf g}_{0} = su(n) \oplus su(n) \oplus u(1)$ and 
${\bf h} = su(n)$ commuting with 
$T = \bar{T} = J_{+} + J_{-}$, 
which are represented respectively by
\be
\left(\begin{array}{ccc}
a & & 0\\
0 & & b \end{array}\right) , ~~ a^{\dagger} = a , ~~ 
b^{\dagger} = b , ~~ {\rm Tr}(a+b) = 0 ~; ~~~~~
\left(\begin{array}{ccc}
a & & 0\\
0 & & a \end{array}\right) , ~~ a^{\dagger} = a , ~~
{\rm Tr} a = 0 ~.
\ee
So the corresponding MIST models are integrable 
perturbations of the WZW coset
\be
{SU(n) \times SU(n) \over SU(n)} \times U(1)
\ee
and the potential term is
\be
{\rm Tr} (g^{-1} T g \bar{T}) = {\rm Tr} 
(g_{1}^{\prime} {g_{2}^{\prime}}^{-1} + 
g_{2}^{\prime} {g_{1}^{\prime}}^{-1})
\ee
by restricting to group elements 
$g^{\prime} = (g_{1}^{\prime} , ~ g_{2}^{\prime})$ in 
$G_{0}^{\prime} = SU(n) \times SU(n)$ when $\chi =0$.
To proceed with the gravitational dressing of the 
resulting integrable models one has to check the
validity of the decoupling condition (19)
in this case. It 
can be easily verified that this condition is 
satisfied only for $n=1$ and $n=2$ since then
${\rm Tr} (g_{1}^{\prime} {g_{2}^{\prime}}^{-1}) 
= {\rm Tr} (g_{2}^{\prime} {g_{1}^{\prime}}^{-1}) 
= {\rm Tr} (g_{1}^{\prime} 
{g_{2}^{\prime}}^{-1})^{-1}$. We will examine each 
case separately.

For $A_{1}$ ($n=1$) the massive integrable soliton theory
with $\chi = 0$ is precisely the sine-Gordon model
\be
\bar{\partial} \partial \theta = 2 m^2 \sin 2 \theta ~,
\ee
where $g_{1}^{\prime} = e^{i \theta}$, 
$g_{2}^{\prime} = e^{-i \theta}$ is a good parametrization
that solves the WZW constraints in (16). If we dress it
gravitationally using the parametrization 
\be
\tilde{g} = \left(\begin{array}{ccc}
e^{i \theta} & & 0\\
0  & & e^{-i \theta} \end{array}\right) 
e^{2 \phi J_{0}} = 
\left(\begin{array}{ccc}
e^{i \theta + \phi} & & 0\\
0 & & e^{-i \theta + \phi} \end{array}\right) ,
\ee
the resulting conformal model is the complexified
Liouville theory whose real and imaginary parts
are respectively
\be
\bar{\partial} \partial \phi = 2 m^2 e^{2 \phi} 
\cos 2 \theta ~; ~~~~~
\bar{\partial} \partial \theta = 2 m^2 e^{2 \phi} 
\sin 2 \theta ~.
\ee

For $A_{3}$ ($n=2$), setting $\chi = 0$, we arrive at the
so-called matrix sine-Gordon model (see [13] for details)
for which the appropriate parametrization of  
$g_{1}^{\prime}$ and $g_{2}^{\prime}$ in $SU(2)$ 
can be found by solving the corresponding WZW constraints
\be
{g_{1}^{\prime}}^{-1} \partial g_{1}^{\prime} + 
{g_{2}^{\prime}}^{-1} \partial g_{2}^{\prime} = 0 ~,
~~~~~ \bar{\partial} g_{1}^{\prime} {g_{1}^{\prime}}^{-1}
+ \bar{\partial} g_{2}^{\prime} {g_{2}^{\prime}}^{-1} 
=0 ~.
\ee
Then, its gravitational dressing proceeds as above to yield
the coupled system of equations
\be
\bar{\partial} ({g_{1}^{\prime}}^{-1} \partial 
g_{1}^{\prime}) = m^2 e^{2 \phi} ({g_{1}^{\prime}}^{-1}
g_{2}^{\prime} - {g_{2}^{\prime}}^{-1} g_{1}^{\prime}) , 
~~~~~ \bar{\partial} \partial \phi = m^2 e^{2 \phi} 
{\rm Tr} ({g_{1}^{\prime}}^{-1} g_{2}^{\prime} + 
{g_{2}^{\prime}}^{-1} g_{1}^{\prime}) ~.
\ee

(ii) {\em C-series}: For $C_{n}$ there is only one $N=1$
embedding of $sl(2)$ specified in the defining  
representation by (25) as before. Then, the zero-graded 
subalgebra is ${\bf g}_{0} = su(n) \oplus u(1)$ and 
${\bf h} = so(n)$, which are represented respectively by
\be
\left(\begin{array}{ccc}
a & & 0\\
0 & & -a^{\star} \end{array} \right) , ~~ 
a^{\dagger} = a ~; ~~~~~ 
\left(\begin{array}{ccc}
a & & 0\\
0 & & a \end{array}\right) , ~~ a^{\star} = -a ~, ~~
a^{t} = -a ~.
\ee
As a result, the corresponding MIST models are integrable
perturbations of 
\be
{SU(n) \over SO(n)} \times U(1) 
\ee
and the potential term takes the form
\be
{\rm Tr} (g^{-1} T g \bar{T}) = {\rm Tr} 
(g^{\prime} {g^{\prime}}^{t} + {g^{\prime}}^{\star} 
{g^{\prime}}^{-1}) ~,
\ee
where the group elements are restricted to 
$(g^{\prime} , ~ {g^{\prime}}^{\star})$ in the $2n$-dim
representation of $G_{0}^{\prime} = SU(n)$ 
(with $g^{\prime} \in SU(n)$) by setting 
$\chi = 0$. It is a matter of straightforward 
calculation to verify that the special condition (19)
is satisfied only for $n=1$ and $n=2$, as in the 
$A$-series above, since only for them we have 
${\rm Tr} (g^{\prime} 
{{g^{\prime}}^{-1}}^{\star}) = {\rm Tr} 
({g^{\prime}}^{\star} {g^{\prime}}^{-1})$. 

The case $n=1$ is trivial reducing to the $A_{1}$ 
sine-Gordon model as before, while
$C_{2}$ provides the only model in this series that is 
eligible to yield a non-Abelian Toda theory upon 
gravitational dressing. The resulting conformal theory
coincides with the gravitationally dressed complex
sine-Gordon, which can also be considered, as we will see
next, as the non-Abelian Toda theory for $B_{2}$.
This is because the vector representation of $C_{2}$
with dimension 4 corresponds to the spinor representation
of $B_{2}$.

(iii) {\em B and D-series}: This is the series of orthogonal
groups $so(n)$, $B_{r}$ if $n= 2r + 1$ and $D_{r}$ if
$n = 2r$, for
which there is an $N=1$ embedding of $sl(2)$ in the
defining $n$-dim representation given by
\be
J_{0} = {i \over 2} \left(\begin{array}{ccc}
0 & 1 & \cdots \\
-1 & 0 & \cdots\\
\vdots & \vdots & \ddots \end{array} \right) , ~~~~~~
J_{+} + J_{-} = i \left(\begin{array}{cccc}
0 & 0 & 0 & \cdots \\
0 & 0 & 1 & \cdots \\
0 & -1 & 0 & \cdots \\
\vdots & \vdots & \vdots & \ddots \end{array} \right) 
\ee
that is purely imaginary and antisymmetric. As a 
consequence ${\bf g}_{0} = so(n-2) \oplus u(1)$ and
${\bf h} = so(n-3)$ commuting with $J_{+} + J_{-}$, 
which are represented respectively by
\be
i \left(\begin{array}{ccc}
0 & a & \cdots \\
-a & 0 & \cdots \\
\vdots & \vdots & b \end{array} \right) , ~~~~~~
i \left(\begin{array}{cccc}
0 & 0 & 0 & \cdots \\
0 & 0 & 0 & \cdots \\
0 & 0 & 0 & \cdots \\
\vdots & \vdots & \vdots & c \end{array}\right) ,
\ee
where $a$ is a real number, $b$ is a real $(n-2)$-dim
antisymmetric matrix and $c$ is a real $(n-3)$-dim 
antisymmetric matrix. The MIST in this case describe
integrable perturbations of the WZW coset
\be
{SO(n-2) \over SO(n-3)} \times U(1)
\ee
and the decoupling condition (19) is always satisfied
for any $n$. Hence, setting $\chi = 0$, which in the 
notation above corresponds to $a= 0$, yields the
WZW coset model $SO(n-2)/SO(n-3)$ plus an integrable 
perturbation that follows from the potential term 
${\rm Tr} (g^{-1} (J_{+} + J_{-}) g (J_{+} + J_{-}))$. 
Then, for all $n$, we obtain the associated $so(n)$ 
non-Abelian Toda theory by coupling to gravity 
according to the general framework.

The simplest non-trivial example is provided by 
$B_{2} = so(5)$.
After solving the WZW constraints (16), one finds
that the massive integrable soliton theory 
for $\chi = 0$ (see [12] for details) 
coincides with the complex sine-Gordon
model for a complex field $u$ satisfying the equation
of motion
\be
\bar{\partial} \partial u + {u^{\star} \partial u 
\bar{\partial} u \over 1 - u u^{\star}} = m^2 u 
(1 - u u^{\star}) ~.
\ee
This model has a residual $U(1)$ symmetry generated 
by $u \rightarrow e^{i \theta} u$ and provides a charged
generalization of the ordinary sine-Gordon model. 
It has been identified with an integrable perturbation
of the $SO(3)/SO(2)$ WZW model, where the potential
term ${\mid u \mid}^2$ represents the first thermal 
parafermion operator in a Lagrangian framework [6, 7].
Setting $u = \sin r \exp (i \theta)$ we obtain a more
standard description of the model in a curved background
geometry with diagonal metric $(1, ~ {\tan}^2 r)$. 
Upon gravitational dressing the resulting conformal
theory is the $B_{2}$ non-Abelian Toda model that was
also considered by Gervais and Saveliev in a different
context [9]. We may introduce the non-compact version
based on the WZW coset $SL(2)/U(1)$ by treating 
$u$ and $u^{\star}$ as the independent variables 
$u = \sinh r \exp (- \theta)$, 
$u^{\star} = - \sinh r \exp (\theta)$. Then, the 
gravitationally dressed model takes the form (6).
It is clear now that for higher values of $n$ we
have multi-component generalizations associated
to the perturbed WZW coset $SO(n-2)/SO(n-3)$ with
the target space field ${\bf u}$ 
(and its conjugate ${\bf u}^{\star}$)
having $n-4$ components. The non-Abelian Toda theory
that arises after gravitational dressing in this case
is summarized by the Lagrangian
\be
L^{\prime} = - {1 \over 1 - {\bf u} {\bf u}^{\star}} 
(\partial {\bf u} \bar{\partial} {\bf u}^{\star} 
+ \bar{\partial} {\bf u} \partial {\bf u}^{\star}) 
+ \partial \phi \bar{\partial} \phi + m^2 e^{2 \phi} 
(1 - 2{\bf u} {\bf u}^{\star}) 
\ee
as it appears in [9]. 

Concluding this class of examples we mention that the
$D$-series admits additional $N=1$ embedding when 
$r$ is an even number $2p$ giving rise to MIST based
on the WZW coset $U(2p)/Sp(p)$. This possibility, as
well as the models based on exceptional algebras will
not be considered here any further. 

It is intriguing, as side comment, that the coset space
geometry of the WZW models $SO(n-2)/SO(n-3)$ describe the
dynamics of the physical degrees of freedom of a bosonic
string propagating in $(n-1)$-dim flat Minkowski space
after solving the classical Virasoro constraints in 
the $X^{0} = \tau$ gauge [14]. In this context the 
potential term ${\mid {\bf u} \mid}^2$ originates from 
self-interactions of the string \`a la Kalb-Ramond, 
as was originally pointed out in four dimensions by 
Lund and Regge [15]. The multi-component generalizations
of the sine-Gordon model based on the WZW coset
$SO(n-2)/SO(n-3)$ arise in yet another context by
performing a conformal reduction \`a la Pohlmeyer
of the ordinary 2-dim non-linear $\sigma$-models 
based on $S^{n-2} = SO(n-1)/SO(n-2)$ [16]; for $n=4$
we obtain the sine-Gordon, for $n=5$ the complex
sine-Gordon, and so on. It will be interesting to
investigate the relevance of the gravitationally 
dressed models (41) to either of these two seemingly 
unrelated frameworks that also give rise to 
generalized  sine-Gordon models. 

One may inquire whether the classically conformal
Toda theories, including their non-Abelian 
generalizations that result by gravitational 
dressing, define quantum mechanically consistent 
string backgrounds. The Toda potential can 
be regarded as an exactly marginal perturbation 
that has been added to the 2-dim $\sigma$-model, 
and as such it plays the role of the tachyon. 
If we consider the corresponding $\beta$-functions 
to lowest order in ${\alpha}^{\prime}$ we have
to find a dilaton field $\Phi$ so that the conformal 
invariance is maintained. Since the 
presence of a tachyon potential $V$ does not affect
$\beta (G_{\mu \nu})$, $\beta (B_{\mu \nu})$ and
$\beta (\Phi )$ to this lowest order, 
we may momentarily neglect $\beta (V)$ and 
examine first whether the non-Abelian
Toda models have all other $\beta$ functions zero.
In their formulation as gravitationally dressed
multi-component sine-Gordon models (c.f. (41)) one
finds that there is no consistent choice of 
dilaton that does the job (see for instance [17]
and references therein). This might seem surprising
bearing their original formulation (24) 
in terms of WZW models. The point is that there are 
two classically equivalent ways to formulate 
these theories, before or after solving the corresponding 
WZW constraints. Choosing the latter in order to 
connect directly with the generalized sine-Gordon 
models, like (40), it unavoidably leads to non-vanishing 
$\beta$-functions. In the other formulation the
$\beta$ function equations are 
satisfied with a linear dilaton $\Phi = Q \phi$, 
but then the tachyon $\beta$-function equation
$\beta (V) = 0$ introduces a non-trivial algebraic
relation between the numerical coefficient $Q$
and $\gamma$ that comes from the 
general conformal factor
$e^{2 \gamma \phi}$; for example, this condition
for the $B_{2}$ non-Abelian Toda theory reads [17]
\be
4 \gamma = Q \pm \sqrt{Q^2 - 
{4 \over {\alpha}^{\prime}}} ~.
\ee
Thus, the quantization of the models in the unreduced 
formulation (24) is more appropriate for string 
applications, although further work is required to
resolve some technical issues that are involved
(analogous to Liouville theory for $c>1$).
For a recent systematic exposition of Toda-like 
$\sigma$-model solutions of string theory see also
[18]. 

A final technical point concerns the type of extended 
conformal symmetries that the models like (41) 
exhibit classically. Recall that the conformal 
coset with Lagrangian 
$(\partial {\bf u} \bar{\partial} {\bf u}^{\star} +
\bar{\partial} {\bf u} \partial {\bf u}^{\star}) 
/(1 - {\bf u} {\bf u}^{\star})$ has a chiral
$W_{\infty}$ symmetry generated by appropriate 
bilinear combinations of the parafermion currents.
Turning on the potential ${\mid {\bf u} \mid}^2$ 
changes these local chiral conservation laws into
non-chiral because the conformal invariance is 
broken but the integrability is preserved (see
[6] for details on these models). The gravitational
dressing restores classically the conformal invariance,
but the curious thing is that the local conservation
laws are not generated by $W_{\infty}$ anymore, 
rather by its consistent truncation restricted only
to elements of even spin. For example, for the $B_{2}$
model there are three chiral currents of spin 2, one
local that is provided by the stress-energy tensor
and a pair of non-local fields, which are essentially 
the gravitationally dressed parafermions [9]. Hence,
going a step further to construct all the local chiral 
conservation laws one has to consider appropriate
bilinears of these parafermions, which yield 
$W$-generators with spin $2, ~4, ~6, \cdots$ in this 
case. This can be verified directly, but we spare
the details of the computation. 
There might be a deeper relation of this
result to the phase diagram of the sine-Gordon model
(and probably its multi-component generalizations), 
where it was found that the overall velocity of the 
renormalization group flow is also cut in half by 
gravity [4, 5]. 

Summarizing, we have found that certain 
multi-component generalizations of the sine-Gordon
model that arise as integrable perturbations 
of WZW cosets away from criticality turn into
conformal non-Abelian Toda theories by coupling
to 2-dim gravity. Although the backgrounds resulting
by imposing classically the WZW constraints 
are not conformal 
quantum mechanically, their integrability should
be preserved in the quantum theory. Our results
offer another justification for focusing 
interest into the theory of non-Abelian Toda 
systems in future investigations, because of their 
relevance to 2-dim gravity; our viewpoint is 
different and perhaps complementary to other 
interpretations that were proposed before for
such non-Abelian models [9]. An interesting question
that arises in this context is the coupling of 
various multi-component sine-Gordon models to 
2-dim gravity using Polyakov's chiral gauge as 
an alternative to the conformal gauge. Finally, 
it will be interesting to examine in detail 
the modifications introduced by 2-dim gravity 
on the renormalization group flow of 
multi-component sine-Gordon models, such as (40),
and study possible gravitational phase transitions
by exhibiting singularities of the corresponding
partition functions. This will provide a systematic
extension of previously known results on the gravitational
sine-Gordon model [4, 5].

\vskip1.5cm
\centerline{\bf Acknowledgements} 
\noindent
We thank H.J. Shin and M.V. Saveliev for useful discussions. 
Q.P. is supported in part by the program of Basic Science 
Research, Korean Ministry of Education BSRI-96-2442, 
and by the Korean Science and Engineering Foundation through
CTP/SNU. He also wishes to thank the hospitality of CTP at
MIT during the final stages of this work.

\newpage
\centerline{\bf REFERENCES}
\begin{enumerate}
\item S. Coleman, ``Dilations", in {\em Aspects of Symmetry:
Selected Erice Lectures}, Cambridge University Press, 1985.
\item A. Zamolodchikov, ``Integrable Field Theory from 
Conformal Field Theory", in Advanced Studies in Pure 
Mathematics \underline{19} (1989) 1.
\item F. David, Mod. Phys. Lett. \underline{A3} (1988) 1651;
J. Distler and H. Kawai, Nucl. Phys. \underline{B321} 
(1989) 509.
\item G. Moore, ``Gravitational Phase Transitions and the 
Sine-Gordon Model", Yale preprint YCTP-P1-92, hep-th/9203061.
\item E. Hsu and D. Kutasov, Nucl. Phys. \underline{B396} 
(1993) 693; C. Schmidhuber, Nucl. Phys. \underline{B404} 
(1993) 342.
\item I. Bakas, Int. J. Mod. Phys. \underline{A9} (1994) 3443.
\item Q-H. Park, Phys. Lett. \underline{B328} (1994) 329; 
Q-H. Park and H.J. Shin, Phys. Lett. \underline{B347} (1995) 
73; ibid \underline{B359} (1995) 125.
\item A. Leznov and M. Saveliev, Lett. Math. Phys. 
\underline{6} (1982) 505; Comm. Math. Phys. \underline{89} 
(1983) 59.
\item J.-L. Gervais and M. Saveliev, Phys. Lett. 
\underline{B286} (1992) 271; A. Bilal, Nucl. Phys. 
\underline{B422} (1994) 258.
\item A. Bilal and I. Kogan, Nucl. Phys. \underline{B449} 
(1995) 569; 
A. Bilal, ``Does Coupling to Gravity Preserve Integrability?",
Ecole Normale preprint LPTENS-96/09, hep-th/9601129.
\item E. Abdalla and M. Abdalla, Phys. Lett. \underline{B365}
(1996) 41.
\item T. Hollowood, J. Miramontes and Q-H. Park, Nucl. Phys. 
\underline{B445} (1995) 451.
\item Q-H. Park and H.J. Shin, Nucl. Phys. \underline{B458}
(1996) 327.
\item I. Bakas and K. Sfetsos, ``Universal Aspects of String
Propagation on Curved Backgrounds", to appear in Phys. Rev. 
D., hep-th/9604195.
\item F. Lund and T. Regge, Phys. Rev. \underline{D14} 
(1976) 1524.
\item I. Bakas, Q-H. Park and H.J. Shin, Phys. Lett. 
\underline{B372} (1996) 45.
\item A. Bilal, ``Consistent String Backgrounds and
Completely Integrable 2-D Field Theories", Ecole Normale
preprint LPTENS-95/32, hep-th/9508062.
\item C. Klimcik and A. Tseytlin, Nucl. Phys. 
\underline{B424} (1994) 71.
\end{enumerate}

\end{document}